\documentclass{elsart_lv}

\usepackage{graphicx,amssymb}
\journal{Chemical Physics Letters}

\def\be{\begin{equation}}
\def\ee{\end{equation}}
\def\bea{\begin{eqnarray}}
\def\eea{\end{eqnarray}}
\newcommand{\ket}[1]{\mbox{$|#1\rangle$}}

\begin{document}

\begin{frontmatter}
\title{NMR Implementation of a Building Block for Scalable Quantum Computation}
\author[P,IBM]{D.E. Chang\thanksref{dcmail}},
\author[EE,IBM]{L.M.K. Vandersypen\corauthref{lvmail}},
\author[EE,IBM]{M. Steffen\thanksref{msmail}}
\thanks[dcmail]{E-mail: dchang@snow.stanford.edu}
\corauth[lvmail]{Corresponding author. E-mail: lieven@snow.stanford.edu}
\thanks[msmail]{E-mail: msteffen@snow.stanford.edu}
\address[P]{Stanford University, Department of Physics, Stanford, CA
94305} 
\address[EE]{Stanford University, Department of Electrical
Engineering, Stanford, CA 94305} \address[IBM]{IBM Almaden Research
Center, San Jose, CA 95120}
\begin{abstract}
We report the implementation of the central building block of the
Schulman-Vazirani procedure for fully polarizing a subset of two-level
quantum systems which are initially only partially polarized. This
procedure consists of a sequence of unitary operations and incurs only
a quasi-linear overhead in the number of quantum systems and
operations required.  The key building block involves three quantum
systems and was implemented on a homonuclear three-spin system using
room temperature liquid state nuclear magnetic resonance (NMR)
techniques.  This work was inspired by the state initialization
challenges in current NMR quantum computers but also shines new light
on polarization transfer in NMR.
\end{abstract}
\end{frontmatter}
%
\section{Introduction}
%
The possibility of constructing computers that operate based on the
laws of quantum physics has been raised by a series of theoretical
results over the last twenty years
\cite{feynman82,deutsch85,shor94}. Compared with classical computers,
the promise of these quantum computers lies in their potential ability
to solve a wider class of problems ``efficiently'' --- that is,
without incurring an overhead in time or physical resources that
increases exponentially with the problem size.  Several recent
experiments brought this theory into practice using liquid-state NMR
techniques, with coupled nuclear spins as quantum bits (qubits). In
these implementations, the spin-up and spin-down states represented
the computational basis states $|0{\rangle}$ and
$|1{\rangle}$~\cite{gershenfeld97,cory97,chuang98,jones98,vandersypen00,jones00}.
This work for the first time demonstrated the feasibility of
small-scale quantum computers.

A characteristic feature of nuclear spin systems in equilibrium at moderate
temperatures is that the spin polarization is typically only about $10^{-5}$
to $10^{-6}$. In other words, the spins are in a highly random or {\em
mixed} initial state, whereas the desired initial state for quantum
computation is a well-known or {\em pure} state.  The ability to perform
quantum computation with a mixed initial state derives from methods to
create {\em effective pure} states~\cite{gershenfeld97,cory97}; these are
mixed states which produce the same signal, up to a scaling factor, as spins
in the corresponding pure state.  The preparation of effective pure states,
however, is inefficient: the resulting signal strength decreases
exponentially with the number of qubits $k$~\cite{gershenfeld97,cory97}.
Clearly, preparing effective pure states in this way is not a scalable
approach to quantum computing.

Very high polarizations could be achieved by cooling the liquid NMR
sample down to the milli-Kelvin regime, but this would freeze the
sample, reintroducing dipolar couplings and resulting in very broad
lines and short coherence times, which is problematic for quantum
computing.  Several proposals exist to address these
complications~\cite{cory00,yamamoto}, but their feasability and
scalability remain to be demonstrated. Other physical cooling methods
could be used to increase the initial polarization. The use of optical
pumping~\cite{fitzgerald98} is promising, but the state of the art (in
liquid state) is still far removed from generating highly polarized
nuclear spins useful for quantum computation. Two proton spins have
recently been hyperpolarized using {\em para} hydrogen and
subsequently used in a quantum computation~\cite{glaser00}, but scaling
to $k$ qubits requires precursor molecules which easily react with
$k/2$ H$_2$ molecules to form a stable quantum computer molecule. It
is unlikely that any of these techniques in themselves will provide
large-scale quantum computers.

Remarkably, an {\em algorithmic} --- and thus technology independent ---
solution to the state initialization problem was recently developed by
Schulman and Vazirani~\cite{schulman99} (a related but less efficient
algorithm was devised earlier by Cleve and DiVincenzo~\cite{cleve96} in the
context of Schumacher compression). Their initialization procedure is based
upon polarization transfer and allows one to fully polarize $k$ qubits
starting from $n$ ($> k$) partially polarized qubits. Polarization transfer
is widely utilized in the NMR community and the bounds on the achievable
polarization enhancements have been well
documented~\cite{sorensen89,sorensen91}. The contribution of Schulman and
Vazirani lies in the development and analysis of an explicit algorithm for
polarization enhancement which asymptotically (for large $k$) approaches the
entropy bound and does so {\em in a runtime quasi-linear in $k$ and using a
total number of spins $n$ which is linear in $k$}. Since the overhead is
only polynomial in $k$ as opposed to exponential in $k$, this algorithm
makes the initialization of room temperature NMR quantum computers, and of
any quantum computer which starts from ``high temperature'' qubits, in
principle scalable.

In this letter, we report the experimental realization of the key step
of the Schulman-Vazirani scheme using liquid state NMR techniques. The
experiment was performed on a homonuclear three-spin system, for which
eigenvalue conservation allows a maximum polarization enhancement
factor of 3/2.

\section{Theoretical bounds on polarization transfer}
\label{sec:bounds}
Let us define the polarization $\epsilon$ of a spin $j$ as the
difference in probabilities between the spin up and spin down state,
tracing out the other spins in the molecule. Mathematically, this is
expressed as $2 \mathrm{Tr}(\tilde{\rho} I_z^j)$, where $\tilde{\rho}$
is the density matrix of the spin system and $I_z^j$ is the usual
angular momentum operator of spin $j$ in the $\hat{z}$ direction. So
spin $j$ is in the ground state $|0\rangle$ with probability
$\frac{1+\epsilon}{2}$ and in the excited state $|1{\rangle}$ with
probability $\frac{1-\epsilon}{2}$.

In thermal equilibrium at room temperature, the polarization
$\epsilon=\epsilon_0$, where $\epsilon_0=\hbar \omega / 2 k T $ is a
very small number, about $10^{-5}$ to $10^{-6}$ for typical magnetic
field strengths, because the thermal energy $kT$ is much larger than
the energy difference $\hbar \omega$ between the spin up and spin down
state.  We will from now on associate a temperature with the
polarization of a spin: a spin with relatively high polarization is
considered ``cold'', while one with low polarization is considered
``warm''.

The polarization of one or more spins can be increased reversibly by
polarization transfer from the remaining spins in the same molecule
via a sequence of RF pulses and delay times.  The maximum polarization
enhancement that can be achieved reversibly is bounded by entropy
considerations. Specifically, all reversible procedures, such as pulse
sequences with negligible relaxation, must conserve the entropy, given
by $H=-\mathrm{Tr}(\tilde{\rho} \, \mbox{log} \,
\tilde{\rho})$. However, bounds established using entropy conservation
are often weak~\cite{sorensen89}, and a stronger universal bound on
spin dynamics must be considered.

The spin dynamics bound on polarization enhancement derives from
eigenvalue conservation of the density matrix, a second necessary
condition for reversibility.  This bound can be calculated via the
deviation density matrix $\rho$, where $\tilde{\rho} = 1/2^n (I +
2 \epsilon_0 \rho)$ (note that the polarization
$2 \mathrm{Tr}(\tilde{\rho} I_z^j)$ is proportional to $2 \mathrm{Tr}
(\rho I_z^j )$, since $I_z$ is traceless). By convention in NMR and
for the sake of simplicity, subsequent use of the term ``density
matrix'' will actually refer to $\rho$.  Suppose that we have an
initial density matrix ${\rho}_i$, and we seek to transform it into
some final density matrix $A$, which may be unachievable when using
only unitary operations.  We then wish to find among all the
achievable final density matrices $\rho_f$ the one that is ``closest''
to $A$ --- that is, whose mathematical projection onto $A$ is
maximized. Equivalently, we wish to maximize the coefficient $a$ in
the following expansion of $\rho_f$:
\be
\label{expansion}
{\rho}_f=aA+bB
\,,
\ee
where $B$ represents a matrix orthogonal to $A$, so that
$\mathrm{Tr}(A^{{\dagger}}B)=0$.  It can be shown \cite{sorensen89}
that
\be
\label{max}
a^{max}=\frac{\mathrm{Tr}({\rho}_i^{D}A^{D})}{\mathrm{Tr}(A^2)}
\,,
\ee
where ${\rho}_i^{D}$ and $A^{D}$ are diagonalized representations of
$A$ and ${\rho}_i$, with the diagonal elements arranged in descending
order from left to right.  When both ${\rho}_i$ and $A$ are already
diagonal, evaluation of (\ref{max}) is particularly simple, reducing
to a rearrangement of the diagonal elements and a trivial trace
calculation.

For a homonuclear three-spin system starting from thermal equilibrium
at room temperature ($\epsilon =\epsilon_0 \ll 1$), the initial
density matrix can be expressed as $\rho_{i}=I_z^{a}+I_z^{b}+I_z^{c}$,
or in explicit matrix form,
\renewcommand{\arraystretch}{0.85}
\renewcommand{\arraycolsep}{1mm}
\be
{\rho}_{i}=\frac{1}{2}
\left( \begin{array}{p{4mm}p{4mm}p{4mm}p{4mm}p{4mm}p{4mm}p{4mm}p{4mm}}
3 & 0 & 0 & 0 & 0 & 0 & 0 & 0 \\
0 & 1 & 0 & 0 & 0 & 0 & 0 & 0 \\
0 & 0 & 1 & 0 & 0 & 0 & 0 & 0 \\
0 & 0 & 0 & -1 & 0 & 0 & 0 & 0 \\
0 & 0 & 0 & 0 & 1 & 0 & 0 & 0 \\
0 & 0 & 0 & 0 & 0 & -1 & 0 & 0 \\
0 & 0 & 0 & 0 & 0 & 0 & -1 & 0 \\
0 & 0 & 0 & 0 & 0 & 0 & 0 & -3
\end{array} \right)
\,.
\ee
If the goal is to maximize the polarization of spin $a$, the desired
final density matrix is $A=I_z^{a}$.  Application of (\ref{max}) then
readily yields a maximum possible projection $a^{max}=3/2$ onto $A$.
Since the projection of $\rho_i$ onto $A$ yields $a^{initial}=1$, the
maximum achievable polarization enhancement factor starting from
thermal equilibrium is $a^{max}/a^{initial}=3/2$.
%
\section{The Schulman-Vazirani Scheme}
%
The bounds on polarization transfer established more than ten years ago
indicate that unitary polarization transfer procedures could in principle
result in very high polarization enhancements~\cite{sorensen89}.  However,
only recently, due to the work of Schulman and Vazirani, did an {\em
explicit and efficient} unitary procedure become available for transforming
the thermal equilibrium state into a state in which a subset of the spins
exhibits a polarization that asymptotically approaches $\epsilon =
1$~\cite{schulman99}.  This procedure was originally invented to initialize
the spins in NMR quantum computing in a scalable way, but it can be used to
efficiently boost the polarization of any subset of a set of partially
polarized qubits, regardless of their physical implementation as long as the
available Hamiltonians can generate the necessary unitary transformations.

The idea behind the Schulman-Vazirani scheme is to redistribute the entropy
over the qubits, such that the entropy of a subset of the qubits approaches
zero while the entropy of the remaining qubits increases and the total
entropy is preserved.  The following ``boosting procedure'' (summarized in
Fig.~\ref{fig:PTcircuit1}) serves as the building block for this
polarization transfer scheme~\cite{schulman99}:

Given three qubits $a,b,$ and $c$ with identical initial polarizations
$\epsilon=\epsilon_0$, the initial state $\ket{x_a} \otimes \ket{x_b}
\otimes \ket{x_c}$, or for short $\ket{x_a} \ket{x_b} \ket{x_c}$, is one of
the eight possible states $\ket{0}\ket{0}\ket{0}, \ket{0}\ket{0}\ket{1},
\ldots,
\ket{1}\ket{1}\ket{1}$, with respective probabilities
$(\frac{1+\epsilon_0}{2})^3$, $(\frac{1+\epsilon_0}{2})^2
(\frac{1-\epsilon_0}{2})$, $\ldots,$ $(\frac{1-\epsilon_0}{2})^3$.\\
First perform a CNOT operation (Table 1) on $c$ conditioned on the
state of $b$.  The new state of the three qubits is $\ket{x_a'}
\ket{x_b'} \ket{x_c'} = \ket{x_a} \ket{x_b} \ket{x_b \oplus x_c}$,
where $\oplus$ denotes addition modulo $2$.  Note that {\em
conditioned} on $\ket{x_c'}=\ket{0}$, the polarization of $b$ is now
$\frac{2\epsilon_0}{1+{\epsilon_0}^2}$ ($b$ is almost twice as cold as
before); {\em conditioned} on $\ket{x_c'}=\ket{1}$, the polarization
of $b$ is $0$ ($b$ is at infinite temperature). However, {\em
overall}, the polarization of $b$ is still the same as before,
$\epsilon_0$. The polarization of $a$ is of course also still
$\epsilon_0$.  We then perform a NOT operation on $c$ followed by a
Fredkin gate (Table 1) with $c$ as the control qubit. The result is
that $a$ and $b$ are swapped if and only if $\ket{x_c'}=\ket{0}$ (and
thus if and only if $b$ has been cooled): $\ket{x_a''} \ket{x_b''}
\ket{x_c''} = \ket{x_b'} \ket{x_a'} \ket{x_c'} $ if
$\ket{x_c'}=\ket{0}$, and $\ket{x_a''} \ket{x_b''} \ket{x_c''} =
\ket{x_a'} \ket{x_b'} \ket{x_c'}$ otherwise. On average, $a$ will thus
be colder than before. The resulting polarization of $a$ is $\epsilon
= \frac{3\epsilon_0}{2}+\mathcal{O}\mathnormal{({\epsilon_0}^3)}$,
where the higher order terms are negligible, so the polarization of
spin $a$ is enhanced by a factor of $3/2$.

In order to achieve increasingly higher polarizations, this boosting
procedure must be applied repeatedly, whereby a fraction of the cold
spins obtained from one round is made progressively colder in the
next.  Spins of little or no polarization are discarded in each
round. Analyzing the polarization transfer using energy and
temperature considerations, Schulman and Vazirani demonstrated
\cite{schulman99} that the progression of rounds can be arranged so
that $k$ bits with nearly optimal enhancement can be extracted.
Specifically, the theoretical maximum $k_{max}$ of zero temperature
$(\epsilon = 1)$ bits that can be extracted is given by the entropy
bound:
\be
k_{max}=(1-H(\epsilon_{0}))n \,,
\ee
where
\be \textstyle
{H(\epsilon)=-\frac{1+\epsilon}{2}\log_{2}\frac{1+
\epsilon}{2}-\frac{1-\epsilon}{2}\log_{2}\frac{1-\epsilon}{2}}
\,,  \ee
and $\epsilon_0$ is the initial polarization. Under the
Schulman-Vazirani scheme (and in agreement with the bounds of
Ref.~\cite{sorensen91}), as $n \rightarrow \infty$, the actual
number of extracted qubits $k \rightarrow k_{max}$, and their
polarization $\epsilon \rightarrow 1$.  Furthermore, the number of
elementary operations (pulses and delay times in NMR) required to
accomplish the entire process is only ${\mathcal{O}} (n \, \mbox{log}
\, n)$.  In summary, the initialization scheme is optimal (it reaches
the entropy bound) and efficient ($k$ is proportional to $n$, and the
runtime is quasi-linear).

\section{Implementation of the boosting procedure}
%
The quantum circuit of Fig.~\ref{fig:PTcircuit1}, which summarizes the
steps in the Schulman-Vazirani boosting procedure, results in the
unitary operation (with $a$ the most significant qubit)
\be
\label{U}
U=
\left( \begin{array}{p{4mm}p{4mm}p{4mm}p{4mm}p{4mm}p{4mm}p{4mm}p{4mm}}
0 & 1 & 0 & 0 & 0 & 0 & 0 & 0 \\
1 & 0 & 0 & 0 & 0 & 0 & 0 & 0 \\
0 & 0 & 1 & 0 & 0 & 0 & 0 & 0 \\
0 & 0 & 0 & 0 & 1 & 0 & 0 & 0 \\
0 & 0 & 0 & 0 & 0 & 1 & 0 & 0 \\
0 & 0 & 0 & 1 & 0 & 0 & 0 & 0 \\
0 & 0 & 0 & 0 & 0 & 0 & 1 & 0 \\
0 & 0 & 0 & 0 & 0 & 0 & 0 & 1
\end{array} \right)
\,,
\ee
which transforms the thermal density matrix as 
\be
I_z^{a}+I_z^{b}+I_z^{c} \,\, {\rightarrow} \,\,
\frac{3}{2} \, I_z^{a} + \frac{1}{2} \, I_z^{b} - I_z^{a}I_z^{c} - I_z^{b}I_z^{c}
\,.
\label{eq:transform}
\ee
The propagator $U$ thus redistributes the populations in such a way
that the highest populations are moved to states with
$|a{\rangle}=|0{\rangle}$.  This can be clearly seen by expressing the
resulting density matrix in explicit matrix form:
\be
{\rho}_{f}=\frac{1}{2}
\left( \begin{array}{p{4mm}p{4mm}p{4mm}p{4mm}p{4mm}p{4mm}p{4mm}p{4mm}}
1 & 0 & 0 & 0 & 0 & 0 & 0 & 0 \\
0 & 3 & 0 & 0 & 0 & 0 & 0 & 0 \\
0 & 0 & 1 & 0 & 0 & 0 & 0 & 0 \\
0 & 0 & 0 & 1 & 0 & 0 & 0 & 0 \\
0 & 0 & 0 & 0 & -1 & 0 & 0 & 0 \\
0 & 0 & 0 & 0 & 0 & -1 & 0 & 0 \\
0 & 0 & 0 & 0 & 0 & 0 & -1 & 0 \\
0 & 0 & 0 & 0 & 0 & 0 & 0 & -3
\end{array} \right)
\,.
\label{eq:rho_f}
\ee

Because the density matrix remains in a diagonal state after
application of each quantum gate in Fig.~\ref{fig:PTcircuit1}, the
boosting procedure can actually be implemented using a simplified
quantum circuit: replacing each gate with a gate whose unitary matrix
is correct up to phases preserves the transformation given by
(\ref{eq:transform}).  Consequently, the Toffoli gate, for which the
fastest known implementation takes on the order of $7/4J$ seconds
(taking all $J_{ij}$ to be $\approx J$), can be substituted with a
Toffoli gate correct up to phases --- consisting of a $90^{\circ}$
$\hat{y}$ rotation of $b$ when $c$ is in $\ket{1}$, followed by a
$180^{\circ}$ $\hat{z}$ rotation of $b$ when $a$ is in $\ket{1}$ and a
$-90^{\circ}$ $\hat{y}$ rotation of $b$ when $c$ is in $\ket{1}$ ---
which takes only $1/J$ seconds.  The actual pulse sequence used in the
experiment is given in Fig.~\ref{pulse}.  This sequence was designed
by standard pulse sequence simplification techniques supplemented by
Bloch-sphere intuition.  The resulting unitary operator is
\be
\tilde{U}=
\left( \begin{array}{p{4mm}p{4mm}p{4mm}p{4mm}p{4mm}p{4mm}p{4mm}p{4mm}}
0 & 1 & 0 & 0 & 0 & 0 & 0 & 0 \\
1 & 0 & 0 & 0 & 0 & 0 & 0 & 0 \\
0 & 0 & -1 & 0 & 0 & 0 & 0 & 0 \\
0 & 0 & 0 & 0 & -1 & 0 & 0 & 0 \\
0 & 0 & 0 & 0 & 0 & 1 & 0 & 0 \\
0 & 0 & 0 & -1 & 0 & 0 & 0 & 0 \\
0 & 0 & 0 & 0 & 0 & 0 & 1 & 0 \\
0 & 0 & 0 & 0 & 0 & 0 & 0 & 1
\end{array} \right)
\,.
\ee

%
\section{Experimental results}
%
A 2 mol $\%$ solution of $\mathrm{C_{2}F_{3}Br}$ in deuterated acetone
was used as the homonuclear three-spin system, for its remarkable
spectral properties: strong chemical shifts (0, 28.2, and 48.1 ppm,
arbitrarily referenced) and large scalar couplings ($J_{ab}=-122.1$
Hz, $J_{ac}=75.0$ Hz, and $J_{bc}=53.8$ Hz) combined with long
relaxation times ($T_2$'s $\approx 4$-$8$ s).  The experiments were
conducted at $30.0^{\circ}$C and $11.7$ Tesla, on a Varian
$\mathrm{^{UNITY}INOVA}$ spectrometer.

All pulses were spin-selective, and varied in duration from $1$ to $3$
ms.  ``Hermite 180'' \cite{warren84} and ``av90'' \cite{abramovich93}
shaped pulses were employed for $180^\circ$ and $90^\circ$ rotations
respectively, in order to minimize the effect of the $J$ couplings
between the selected and non-selected spins during the
pulses. Couplings between the unselected spins are irrelevant whenever
those spins are along $\pm \hat{z}$.
Bloch-Siegert shifts~\cite{emsley90} were accounted for in the pulse
sequence out of necessity: they result in an extra phase acquired by the
non-selected spins in their respective on-resonance reference frames. These
phase shifts are in some cases more than $90^\circ$ per pulse, while even
phase shifts on the order of $5^\circ$ are unacceptable.  For simplicity,
Bloch-Siegert corrections and other $\hat{z}$ rotations were implicitly
performed by changing the phase of subsequent RF pulses.
The duration of the entire sequence of Fig.~\ref{fig:PTcircuit1}
is about $70$ ms.

The theoretical predictions for the spectrum of each spin after the boosting
procedure can be derived most easily from (\ref{eq:rho_f}), taking into
account the sign and magnitude of the $J$-couplings. After a readout pulse
on spin $a$, the four spectral lines in the spectrum of $a$ should ideally
have normalized amplitudes $1:2:1:2$, compared to $1:1:1:1$ for the thermal
equilibrium spectrum (for spins $b$ and $c$, the boosting procedure ideally
results in normalized amplitudes of $0:1:0:1$ and $-1:0:0:1$, respectively).
So the prediction is that the boosting procedure increases the signal of
spin $a$ averaged over the four spectral lines by a factor of 3/2, equal to
the bound for polarization enhancement established in
Section~\ref{sec:bounds}.

The experimentally measured spectra before and after the boosting procedure
are shown in Fig.~\ref{readout}.  Clearly, the signal of spin $a$ has
increased on average as a result of the boosting procedure, and the relative
amplitudes of the four lines are in excellent agreement with the theoretical
predictions. The measured areas under the four peaks combined before and
after polarization transfer have a ratio of $1.255{\pm}0.002$.  The spectra
of spins $b$ and $c$ after the boosting procedure are also in excellent
agreement with the theoretical predictions, up to a small overall reduction
in the signal strength. The proper operation of the boosting procedure is
further validated via the experimentally measured density matrix
(Fig.~\ref{dm}), which demonstrates not only that the boosting procedure
exchanges the populations as intended, but also that it doesn't
significantly excite any coherences.  The experimentally measured
$\mathrm{Tr}({\rho}_{f}I_z^a)/\mathrm{Tr}({\rho}_{i}I_z^a)$ gives a
polarization enhancement factor of $1.235{\pm}0.016$, consistent with the
enhancement obtained just from the peak integrals of spin $a$. The
experimental implementation of the boosting procedure thus successfully
increased the polarization of spin $a$.

Despite the excellent qualitative agreement between the measured and
predicted data, the quantitative polarization enhancement of spin $a$
is lower than ideally achievable.  Given the absence of substantial
coherences (Fig.~\ref{dm}), we attribute this suboptimal enhancement
primarily to signal attenuation due to RF field inhomogeneity and, to
a lesser extent, due to transverse relaxation. The minor excitation of
coherences is attributed mostly to incomplete removal of undesired
coupled evolution during the RF pulses.
%
\section{Summary}
%
We have experimentally demonstrated the building block for the
hyperpolarization procedure outlined by Schulman and Vazirani on a
homonuclear three-spin system.  However, the repeated boosting
required in a much larger spin system would be counteracted by
relaxation and other causes of signal decay, such as RF field
inhomogeneity. Also, it should be noted that when starting from
thermal equilibrium at room temperature, the overhead in the number of
nuclear spins required for the complete Schulman-Vazirani scheme is
enormous~\cite{schulman99}, despite its linear scaling: with
$\epsilon_0 \, {\approx} \, 3{\times}10^{-5}$, at most one out of
every $10^{9}$ spins in a molecule can be fully polarized.  Yet, the
Schulman-Vazirani scheme could become practical for enabling scalable
quantum computation in conjunction with other techniques which
increase the initial polarization, or in other quantum bit
implementations with higher initial polarizations.

Our implementation and discussion of the Schulman-Vazirani scheme also
shines new light on polarization transfer bounds and techniques in
NMR. The polarization of one of three spins was boosted using an
explicit and general protocol for achieving optimal polarization
transfer in arbitrary spin systems. By analysis of energy and
temperature limits, it has been shown that this protocol asymptotically
achieves the entropy bound as $k \rightarrow \infty$ and does so {\em
efficiently}, i.e. without incurring any exponentially growing
overhead.

\section{Acknowledgements}

We thank C.S. Yannoni for preparing the sample, I.L. Chuang, O.W.
S{\o}rensen, and T. Schulte-Herbrueggen for useful discussions,
X. Zhou for helpful comments to the manuscript and W. Risk and
J.S. Harris for support.  LV gratefully acknowledges a Yansouni Family
Stanford Graduate Fellowship.  This work was supported by DARPA under
the NMRQC Initiative.

\clearpage

\vspace*{3cm}
\begin{table}
\begin{center}
\begin{tabular}{|cc|cc|cc|cc|} \hline
\multicolumn{2}{|c|}{\bfseries NOT} 
& \multicolumn{2}{c|}{\bfseries CNOT} 
& \multicolumn{2}{c|}{\bfseries Fredkin} 
& \multicolumn{2}{c|}{\bfseries Toffoli} \\ \hline
in & out & in & out & in & out & in & out \\
0 & 1 & 00 & 00 & 000 & 000 & 000 & 000 \\
1 & 0 & 01 & 01 & 001 & 001 & 001 & 001 \\
  &   & 10 & 11 & 010 & 010 & 010 & 010 \\
  &   & 11 & 10 & 011 & 011 & 011 & 011 \\
  &   &    &    & 100 & 100 & 100 & 100 \\
  &   &    &    & 101 & 110 & 101 & 101 \\
  &   &    &    & 110 & 101 & 110 & 111 \\
  &   &    &    & 111 & 111 & 111 & 110 \\
\hline
\end{tabular}
\end{center}
\bigskip
Table 1: Truth tables for the NOT, controlled-not (CNOT),
controlled-swap (Fredkin), and doubly controlled-not (Toffoli)
operations.  The NOT operation flips an individual qubit.  The CNOT
gate flips the ``target'' qubit if the control qubit is $1$.  The
Fredkin gate swaps two qubits conditional on the control qubit being
$1$.  In this table, it is assumed that the first qubit represents the
control qubit in the CNOT and Fredkin gates.  The Toffoli operation
flips the target qubit conditional on two other qubits being $1$.  It
is assumed in this table that the Toffoli is conditional upon the
first two qubits being $1$.
\label{table:truth}
\end{table}

\clearpage

\begin{center}
FIGURE CAPTIONS
\end{center}

\vspace*{1cm}

\begin{figure}[h]
\caption{A quantum circuit that implements the boosting procedure. The
qubits are represented by horizontal lines and their states are
transformed by circuit elements representing unitary
operations. The $\oplus$ symbol indicates a bit flip.  The $\bullet$
symbol denotes a control qubit --- the operation it controls is performed if and only if
the control qubit is in the state $\ket{1}$.  The controlled-swap operation has been
replaced by an equivalent set of gates: two CNOT's and a Toffoli (Table 1).}
\label{fig:PTcircuit1}
\end{figure}

\vspace*{1cm}

\begin{figure}[h]
\caption{Pulse sequence to implement the boosting procedure.  $X$,
$Y$, and $Z$ represent $90^{\circ}$ rotations about those respective
axes. $\bar{X}$ represents a negative $90^{\circ}$ rotation,  $X^2$
denotes a $180^{\circ}$ rotation and $X^{1/2}$ denotes a $45^\circ$
rotation.  This pulse sequence is designed for molecules with
$J_{ab}<0$ and $J_{ac}, J_{bc}>0$.}
\label{pulse}
\end{figure}

\vspace*{1cm}

\begin{figure}[h]
\caption{Experimentally measured spectra of spin $a$ (Left), spin $b$
(Center) and spin $c$ (Right), after a readout pulse on the corresponding
spin, for the spin system in thermal equilibrium (Top) and after applying
the boosting procedure (Bottom). The real part of the spectra is shown, and
the spectra were rescaled in order to obtain unit amplitude for the thermal
equilibrium spectra. Frequencies are in Hz with respect to the Larmour
frequency of the respective spins.}
\label{readout}
\end{figure}

\vspace*{1cm}

\begin{figure}[h]
\caption{Pictorial representation of the theoretical (left) and experimentally 
measured (right) density matrices, shown in magnitude with the sign of the 
real part (all imaginary components were very small).}
\label{dm}
\end{figure}

\clearpage

\vspace*{1.5cm}

\begin{center}
\includegraphics*[width=4cm]{PTcircuit1.epsf}
\end{center}

\begin{center}
Figure 1
\end{center}
\vspace*{4cm}

\begin{center}
\includegraphics*[width=14cm]{sequence_negativeJab.epsf}
\end{center}

\begin{center}
Figure 2
\end{center}

\clearpage

\begin{center}
\includegraphics*[angle=0,width=4.2cm]{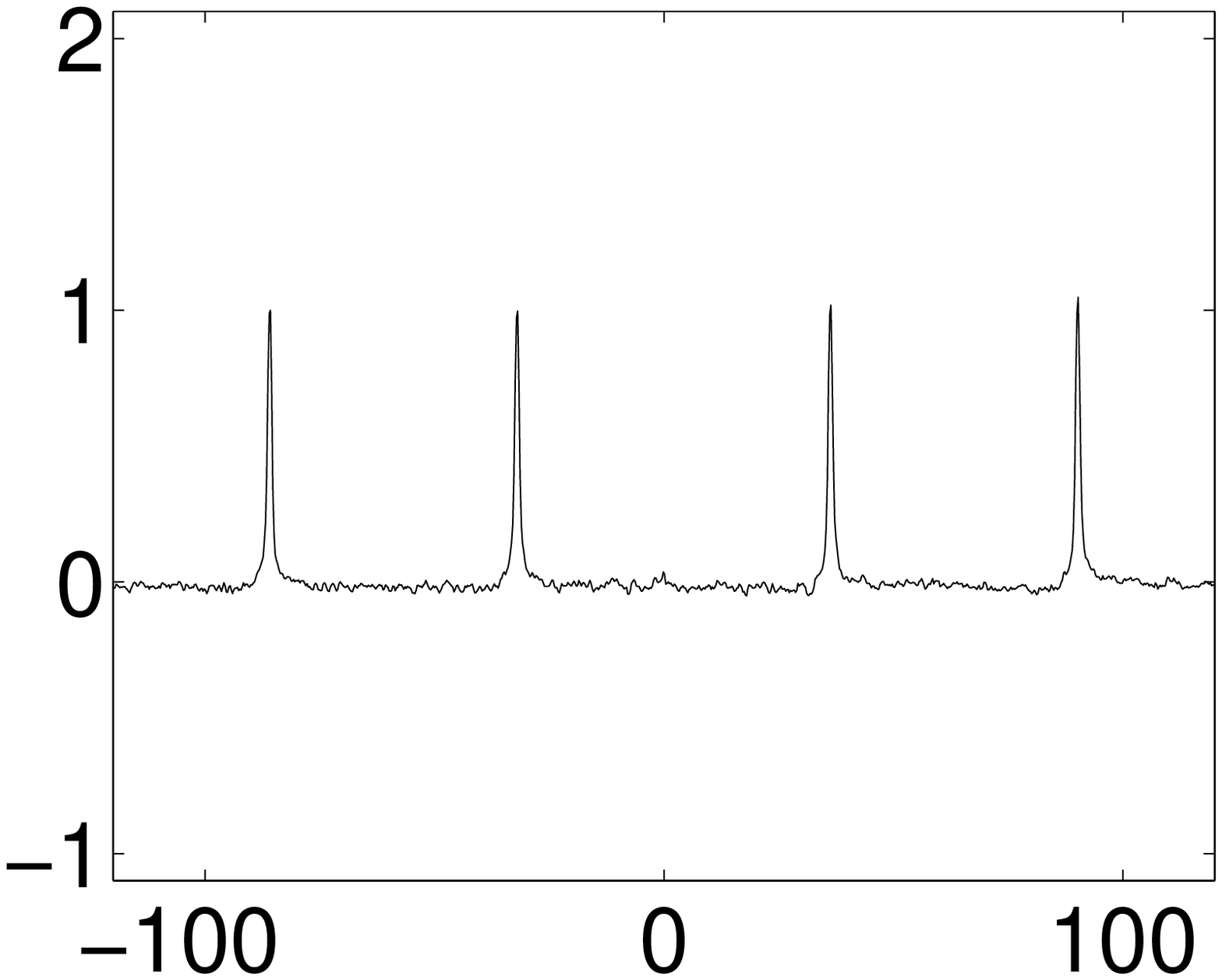} \hspace*{-1ex}
\includegraphics*[angle=0,width=4.2cm]{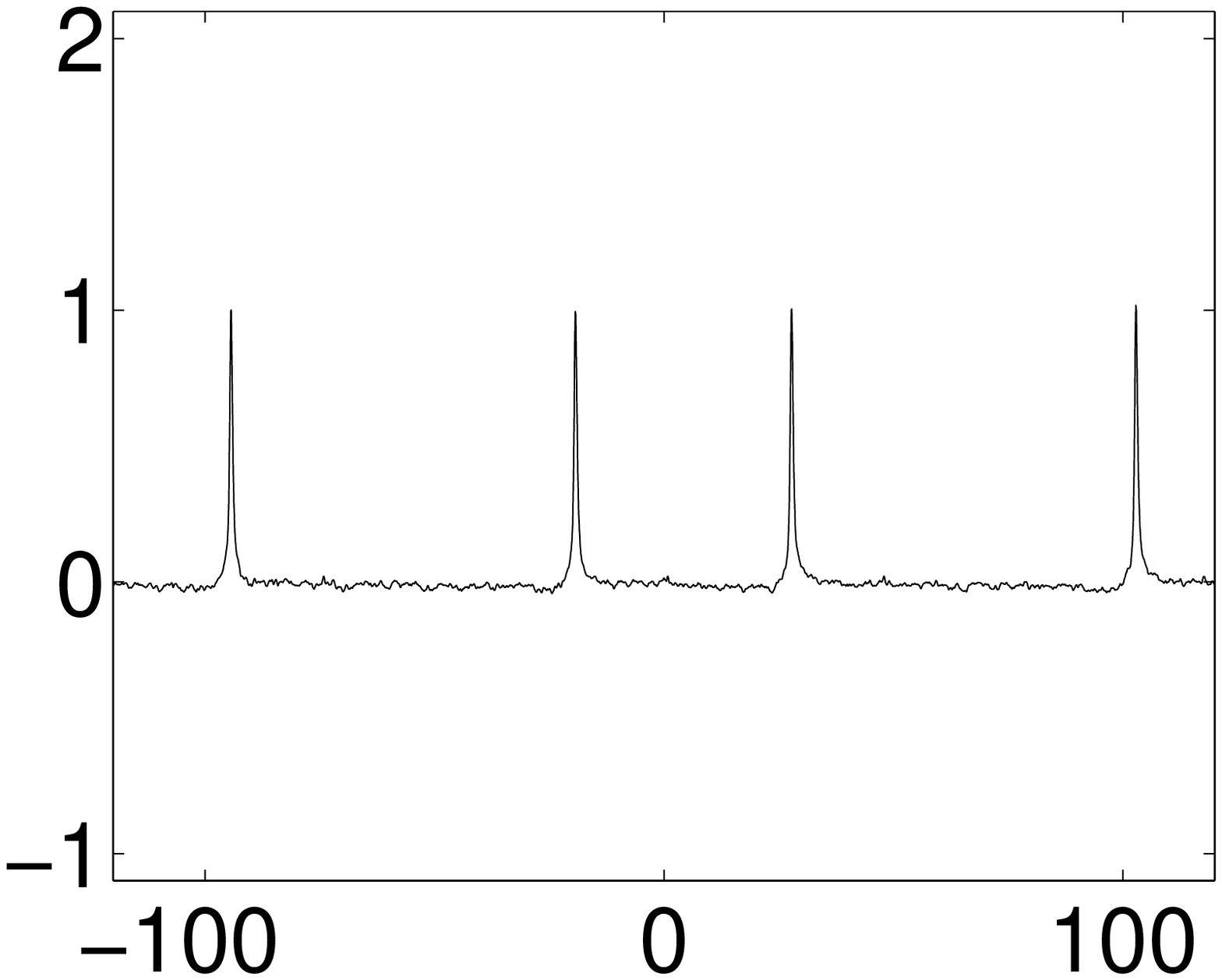} \hspace*{1ex}
\includegraphics*[angle=0,width=4.2cm]{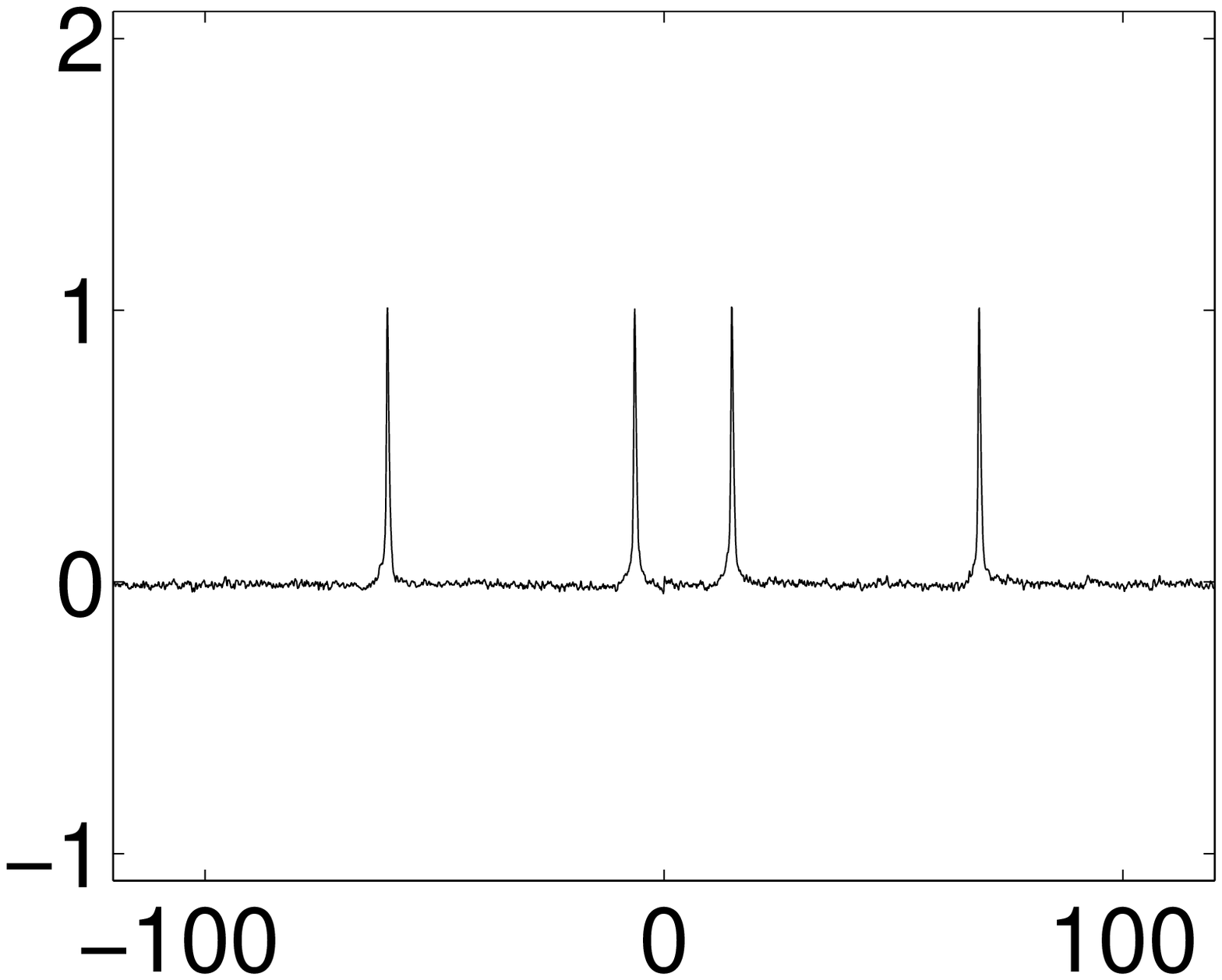}
\includegraphics*[angle=0,width=4.2cm]{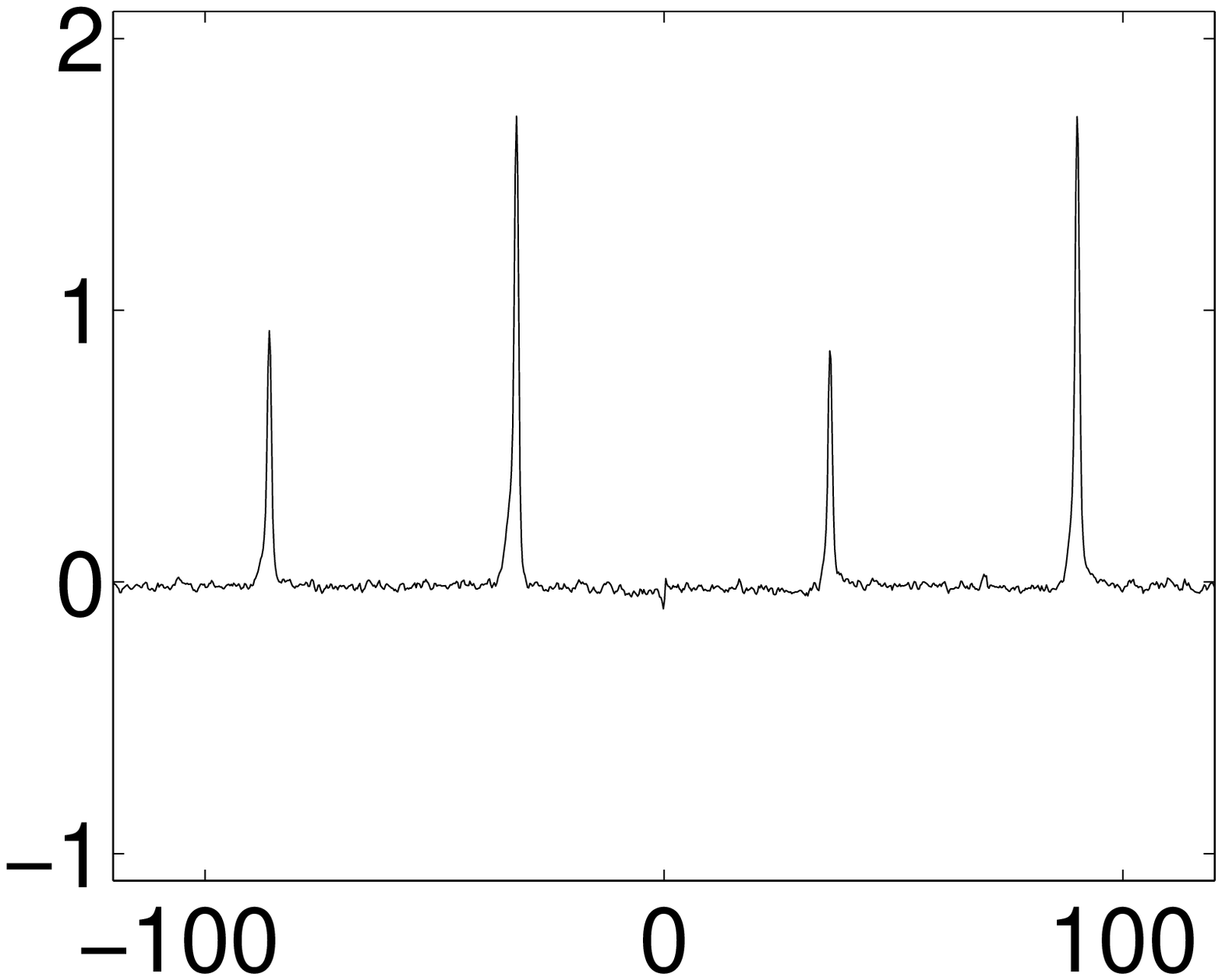} \hspace*{-1ex}
\includegraphics*[angle=0,width=4.2cm]{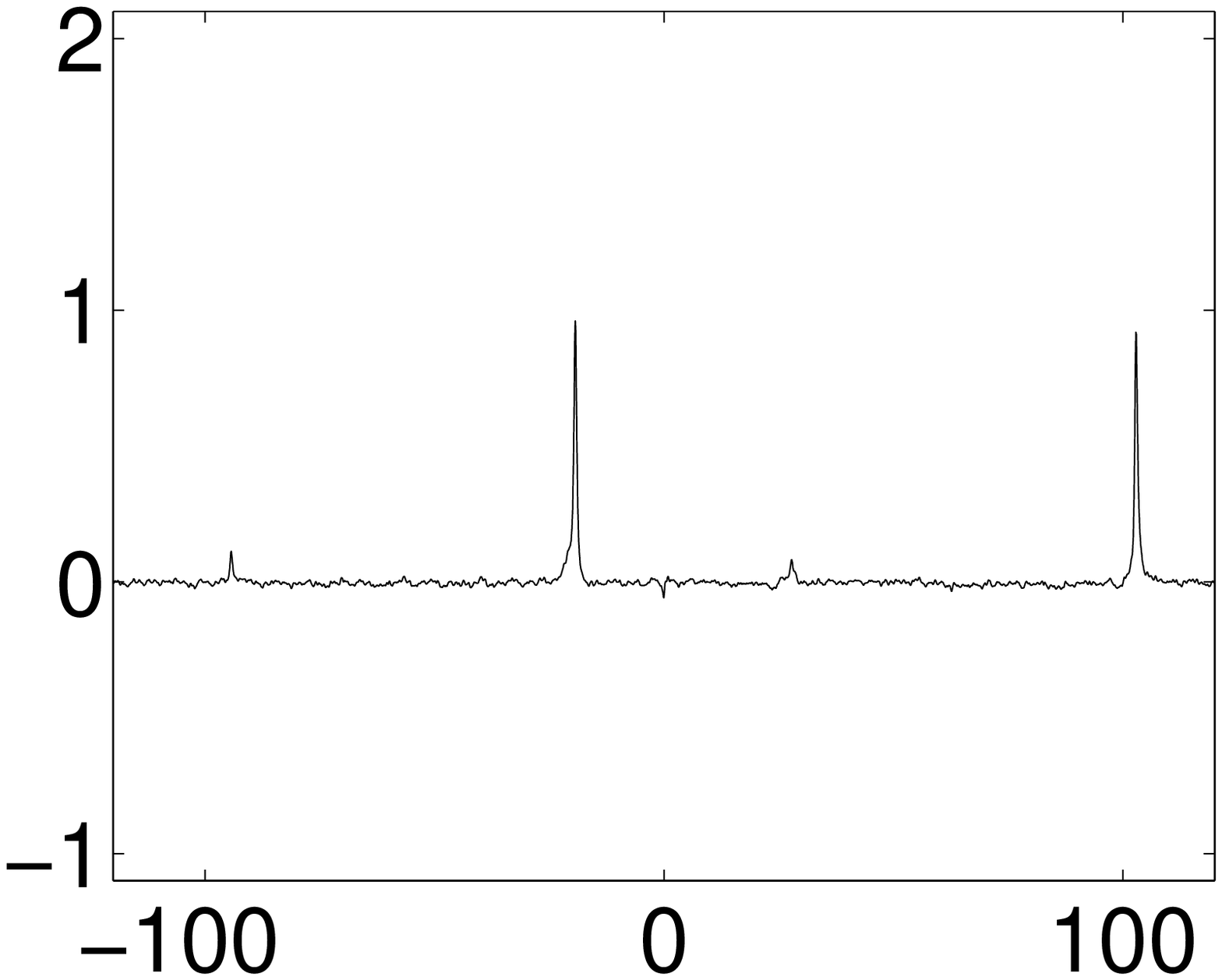} \hspace*{1ex}
\includegraphics*[angle=0,width=4.2cm]{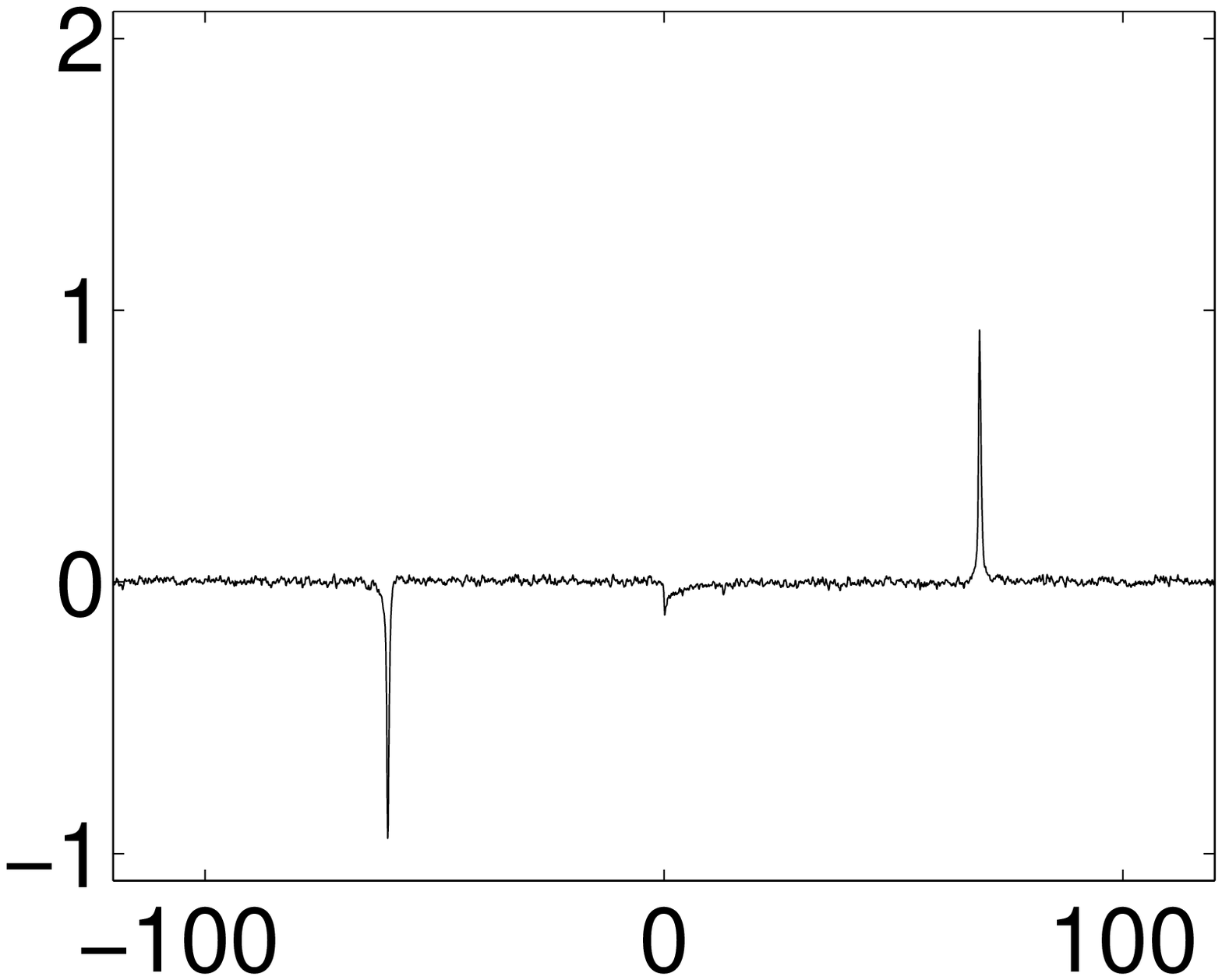}
\end{center}
\vspace*{1cm}

\begin{center}
Figure 3
\end{center}
\vspace*{3cm}

\begin{center}
\includegraphics*[width=6cm]{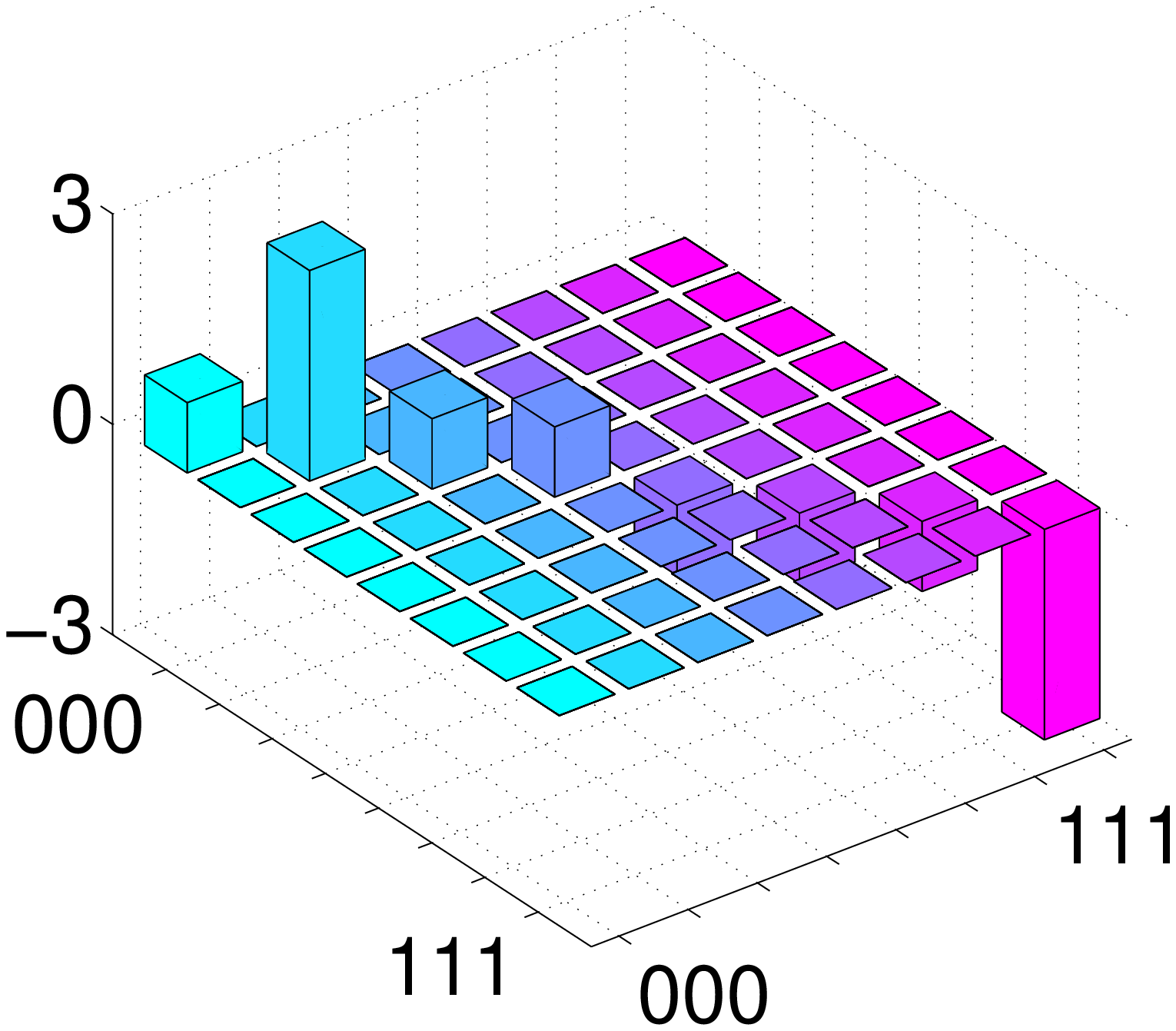}
\includegraphics*[width=6cm]{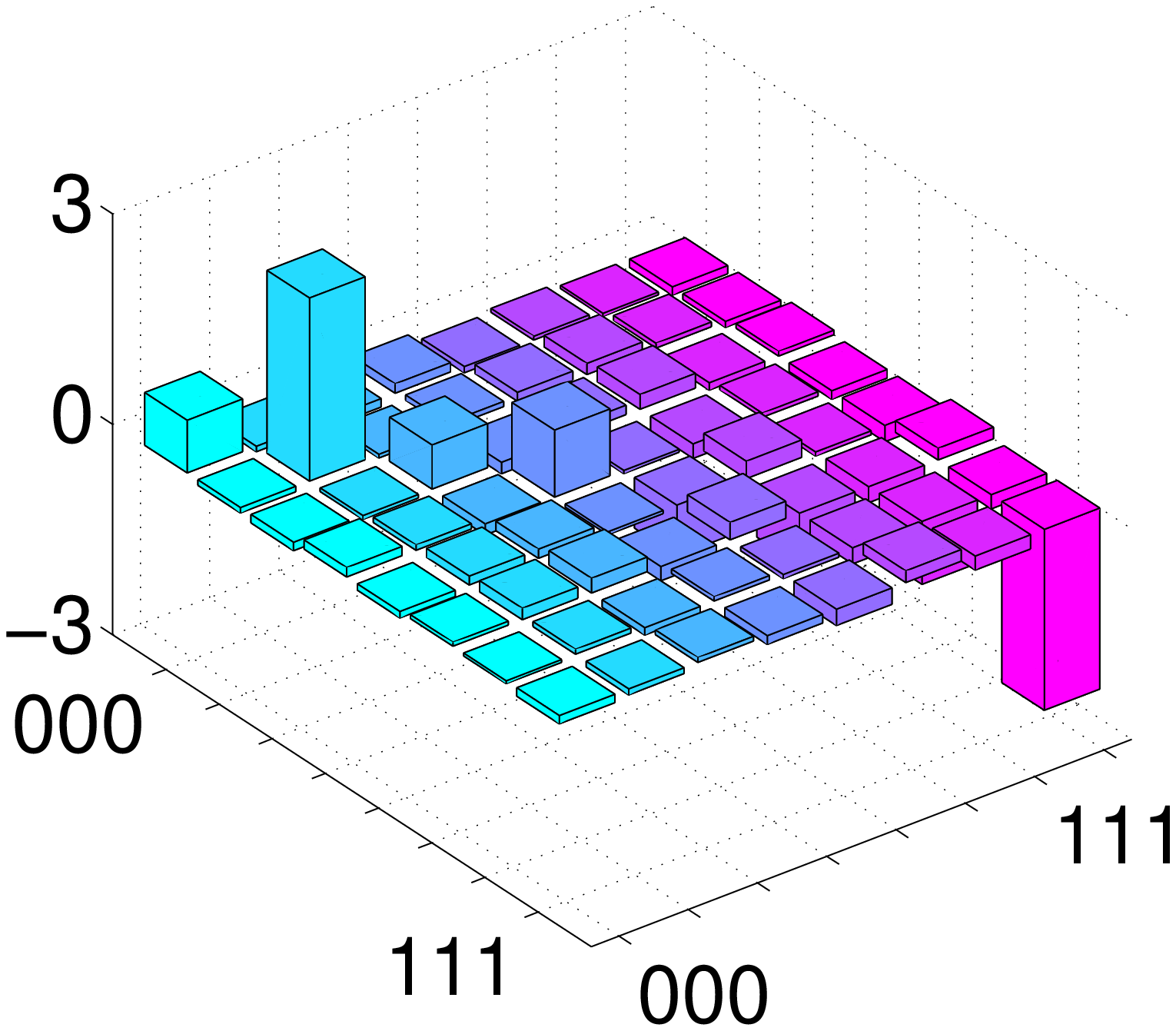}
\end{center}
\vspace*{1cm}

\begin{center}
Figure 4
\end{center}

\end{document}